\documentclass[sigconf]{acmart}
\usepackage{booktabs} 

\copyrightyear{2019}
\acmYear{2019}
\setcopyright{rightsretained}
\acmConference[PEARC '19]{Practice and Experience in Advanced Research Computing}{July 28-August 1, 2019}{Chicago, IL, USA}
\acmBooktitle{Practice and Experience in Advanced Research Computing (PEARC '19), July 28-August 1, 2019, Chicago, IL, USA}
\acmDOI{10.1145/3332186.3333258}
\acmISBN{978-1-4503-7227-5/19/07}

\begin{document}
\title{SciTokens: Demonstrating Capability-Based Access to Remote Scientific Data using HTCondor}

\author{Alex Withers}
\affiliation{%
  \institution{NCSA}
}
\email{alexw1@illinois.edu}

\author{Brian Bockelman}
\affiliation{%
  \institution{Morgridge Institute for Research}
}
\email{bbockelman@morgridge.org}

\author{Derek Weitzel}
\affiliation{%
  \institution{University of Nebraska-Lincoln}
}
\email{dweitzel@cse.unl.edu}

\author{Duncan Brown}
\affiliation{%
  \institution{Syracuse University}
}
\email{dabrown@syr.edu}

\author{Jason Patton}
\affiliation{%
    \institution{University of Wisconsin-Madison}
}
\email{jpatton@cs.wisc.edu}

\author{Jeff Gaynor}
\affiliation{%
  \institution{NCSA}
}
\email{gaynor@illinois.edu}

\author{Jim Basney}
\orcid{0000-0002-0139-0640}
\affiliation{%
  \institution{NCSA}
}
\email{jbasney@illinois.edu}

\author{Todd Tannenbaum}
\affiliation{%
  \institution{University of Wisconsin-Madison}
}
\email{tannenba@cs.wisc.edu}

\author{You Alex Gao}
\affiliation{%
  \institution{University of Illinois}
}
\email{yougao2@illinois.edu}

\author{Zach Miller}
\affiliation{%
  \institution{University of Wisconsin-Madison}
}
\email{zmiller@cs.wisc.edu}

\renewcommand{\shortauthors}{A. Withers et al.}
\renewcommand{\shorttitle}{SciTokens}

\begin{abstract}
The management of security credentials (e.g., passwords, secret keys) for computational science workflows is a burden for scientists and information security officers. Problems with credentials (e.g., expiration, privilege mismatch) cause workflows to fail to fetch needed input data or store valuable scientific results, distracting scientists from their research by requiring them to diagnose the problems, re-run their computations, and wait longer for their results. SciTokens introduces a capabilities-based authorization infrastructure for distributed scientific computing, to help scientists manage their security credentials more reliably and securely. SciTokens uses IETF-standard OAuth JSON Web Tokens for capability-based secure access to remote scientific data. These access tokens convey the specific authorizations needed by the workflows, rather than general-purpose authentication impersonation credentials, to address the risks of scientific workflows running on distributed infrastructure including NSF resources (e.g., LIGO Data Grid, Open Science Grid, XSEDE) and public clouds (e.g., Amazon Web Services, Google Cloud, Microsoft Azure). By improving the interoperability and security of scientific workflows, SciTokens 1) enables use of distributed computing for scientific domains that require greater data protection and 2) enables use of more widely distributed computing resources by reducing the risk of credential abuse on remote systems.

In this extended abstract, we present the results over the past year of our open source implementation of the SciTokens model and its deployment in the Open Science Grid, including new OAuth support added in the HTCondor 8.8 release series.
\end{abstract}

%
%

\ccsdesc[500]{Security and privacy~Authorization}

\keywords{OAuth, capabilities, distributed computing}

\settopmatter{printfolios=true}
\maketitle

\section{Introduction}

We introduced the SciTokens model last year in \cite{SciTokensPEARC18}. In this extended abstract, we present the implementation and deployment experience we have gained since then, including new OAuth support added in the HTCondor 8.8 release series.

SciTokens applies the well-known principle of capability-based security to remote data access. Rather than sending unconstrained identity tokens with compute jobs, we send capability-based access tokens. These access tokens grant the specific data access rights needed by the jobs, limiting exposure to abuse. These tokens comply with the IETF OAuth standard \cite{RFC6749}, enabling interoperability with the many public cloud storage and computing services that have adopted this standard. By improving the interoperability and security of scientific workflows, we 1) enable use of distributed computing for scientific domains that require greater data protection and 2) enable use of more widely distributed computing resources by reducing the risk of credential abuse on remote systems.

\begin{figure}
\includegraphics[scale=0.25]{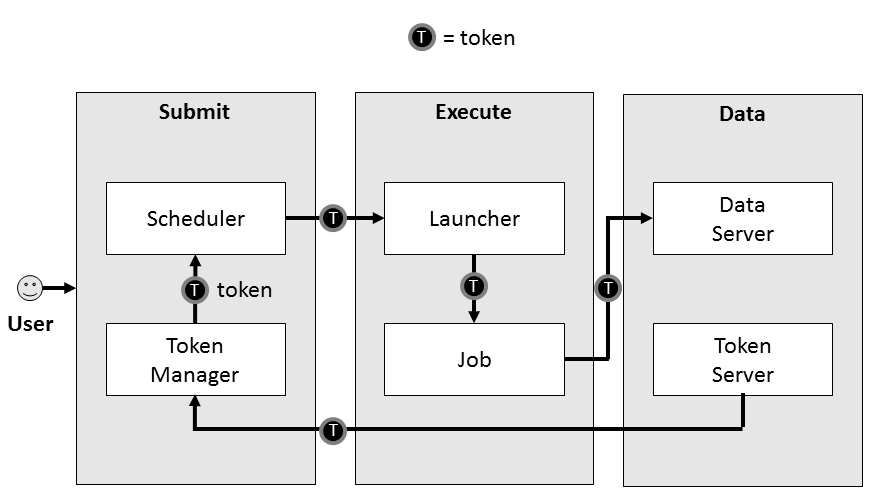}
\caption{The SciTokens Model}
\label{fig:model}
\end{figure}

As illustrated in Figure~\ref{fig:model} from \cite{SciTokensPEARC18}, our SciTokens model applies capability-based security to three common domains in the computational science environment: Submit (where the researcher submits and manages scientific workflows), Execute (where the computational jobs run), and Data (where remote read/write access to scientific data is provided). The Submit domain obtains the needed access tokens for the researcher's jobs and forwards the tokens to the jobs when they run, so the jobs can perform the needed remote data access. The Scheduler and Token Manager work together in the Submit domain to ensure that running jobs have the tokens they need (e.g., by refreshing tokens when they expire) and handle any errors (e.g., by putting jobs on hold until needed access tokens are acquired). The Data domain contains Token Servers that issue access tokens for access to Data Servers. Thus, there is a strong policy and trust relationship between Token Servers and Data Servers. In the Execute domain, the job Launcher delivers access tokens to the job's environment, enabling it to access remote data.

\begin{figure}
\includegraphics[scale=0.25]{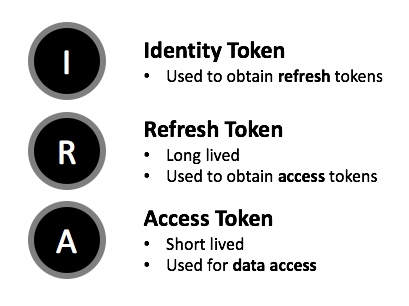}
\caption{Different Token Types}
\label{fig:token_types}
\end{figure}

The SciTokens model adopts token types from OAuth (see Figure~\ref{fig:token_types}). Users authenticate with identity tokens to submit jobs (workflows), but identity tokens do not travel along with the jobs. Instead, at job submission time the Token Manager obtains OAuth refresh tokens with needed data access privileges from Token Servers. The Token Manager securely stores these relatively long-lived refresh tokens locally, then uses them to obtain short-lived access tokens from the Token Server when needed (e.g., when jobs start or when access tokens for running jobs near expiration). The Scheduler then sends the short-lived access tokens to the jobs, which the jobs use to access remote data. 

The remainder of our extended abstract is organized as follows. In Section \ref{sec:htcondor}, we provide an overview of HTCondor's support for SciTokens. In Section \ref{sec:htcondorupdates}, we describe recent HTCondor updates related to SciTokens. In Section \ref{sec:osg}, we describe our Open Science Grid deployment experience. In Section \ref{sec:c}, we describe our new C/C++ implementation. In Section \ref{sec:oauth}, we describe our OAuth server updates. Lastly, in Section \ref{sec:relatedwork} we describe our interoperability efforts with related work, and we conclude in Section \ref{sec:conclusions}.

\section{\label{sec:htcondor}HTCondor Implementation}

As the component that actually executes a scientific workflow, HTCondor serves as the linchpin that ties together all the SciTokens components. To best communicate our HTCondor approach, we first present a walk-thru of how HTCondor orchestrates the component interactions upon submission of a job, followed by a discussion of integration points.

\begin{figure}
\includegraphics[scale=0.25]{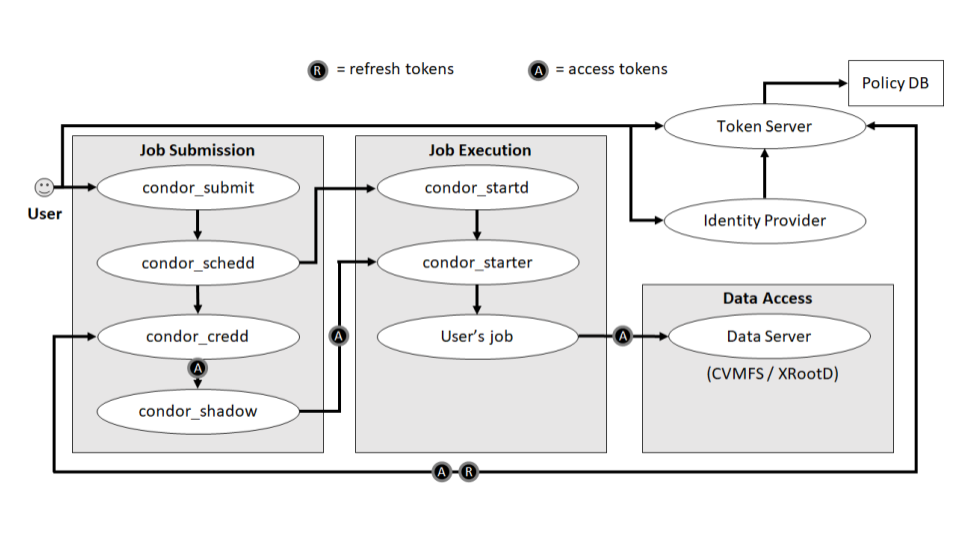}
\caption{The SciTokens System Architecture}
\label{fig:arch}
\end{figure}

As illustrated in Figure~\ref{fig:arch} from \cite{SciTokensPEARC18}, the process begins when the researcher submits the computational job using the condor\_submit command (or more likely using Pegasus or similar workflow front-end that then runs condor\_submit). As part of the submission, the researcher specifies required scientific input data and locations for output data storage in the condor\_submit input file. For example, in a LIGO PyCBC \cite{PyCBC16} submission, the researcher will specify a set of data "frames" from the LIGO instrument that are the subject of the analysis. Then condor\_submit authenticates the researcher to the token\_server(s) to obtain the tokens needed for the job's data access; as an optimization, condor\_submit may first check for any locally cached tokens from the researcher's prior job submissions. The token\_server determines if the researcher is authorized for the requested data access, based on the researcher's identity and/or group memberships or other researcher attributes. If the authorization check succeeds, the Token Server issues an OAuth refresh token back to condor\_submit, which stores the refresh token securely in the condor\_credd, and sends the job information to the condor\_schedd. Since condor\_submit gathers all the needed data access tokens, there is no need to store any identity credentials (e.g., passwords, X.509 certificates, etc.) with the job submission, thereby achieving our goal of a capability-based approach.

The next phase of the process begins when the condor\_schedd has scheduled the job on a remote execution site. The condor\_schedd communicates with the condor\_startd to launch the job, establishing a secure communication channel between the condor\_shadow on the submission side and the condor\_starter on the remote execution side. The condor\_starter then requests access tokens from the condor\_shadow for the job's input data. The condor\_shadow forwards the access token requests to the condor\_credd, which uses its stored refresh tokens to obtain fresh access tokens from the token\_server. The condor\_credd returns the access tokens to the condor\_shadow which forwards them securely to the condor\_starter which provides them to the researcher's job. Note that only access tokens are sent to the remote execution environment; the longer-lived refresh tokens remain secured in the submission environment which typically resides at the researcher's home institution. Lastly, the job uses the access tokens to mount CVMFS filesystem(s) to access scientific data. CVMFS verifies each access token to confirm that the token was issued by its trusted token\_server and that the token's scope includes access to the scientific data being requested. If verification succeeds, CVMFS grants the requested data access. If the access token needs to be refreshed, the condor\_starter makes another request back to the condor\_shadow.

Note that SciTokens can leverage additional aspects of HTCondor, such as the fact that the condor\_shadow can be made explicitly aware if the job is staging input data, accessing data online while the job is running, or staging output data. We allow the job submission to state three different sets of access tokens, which will only be instantiated at file stage-in, execution, and file stage-out, respectively. This enables long running jobs, for instance, to fetch a very short-lived write token for output that will only be instantiated once processing has completed. We also adjust the granularity of access token restrictions; for instance, the condor\_shadow may request fresh access tokens for each job instance, allowing the token to be restricted in origin to a specific execution node. Alternatively, for greater scalability, access tokens can be cached at the credd and shared across all condor\_shadow processes serving jobs that need the same data sets. Finally, we are investigating scenarios in which the data service and its accompanying token service is not fixed infrastructure, but instead is dynamically deployed upon execute nodes, perhaps by the workflow itself. In this scenario, the token service could be instantiated with a set of recognized refresh tokens a priori.

\section{\label{sec:htcondorupdates}HTCondor Updates}

In the past year our OAuth-enabled CredMon progressed from a research prototype to a supported component of HTCondor 8.8 \cite{HTCondor882}. The CredMon is a plug-in to the condor\_credd that implements support for OAuth credentials. Since we use the OAuth standard, it was relatively straightforward to support both SciTokens and Box.com credentials in the same CredMon, so HTCondor jobs can access files both in CVMFS (using SciTokens) and in Box.com folders (using Box tokens).

OAuth support in HTCondor requires a lightweight web server on the submission node, to enable users to authorize credential issuance from the token server (e.g., SciTokens or Box.com) to the condor\_credd. Packaging and configuring this web server component for successful deployment by HTCondor users was a source of multiple lessons learned over the past year. Of special note is the challenge of transferring credentials from the web application to the condor\_credd, which are running under different service accounts for proper isolation. When the OAuth protocol delivers temporary credentials (the OAuth "code") to the web application, the web application writes the credentials to a roundevous directory that the condor\_credd can read from to obtain longer-lived refresh tokens and access tokens. Thus, the more sensitive credentials are not exposed to the web application.

\section{\label{sec:osg}OSG Deployment Experience}

Open Science Grid (OSG) has been an early adopter of the SciTokens model. Over the past year, 13 OSG users have used SciTokens credentials to secure almost two million StashCP uploads across over two thousand servers at 60 unique sites.

OSG currently uses SciTokens with HTCondor in "Local CredMon Mode". In contrast to HTCondor's "OAuth CredMon Mode", the "Local CredMon Mode" configuration uses a SciTokens credential issuer that's local to the HTCondor submit node, that issues credentials according to project-specific policies set by the submit node administrator. Since this mode does not use OAuth, the submitter does not see an OAuth consent screen, but instead HTCondor transparently adds the needed SciToken credentials to the user's job environment to enable project-specific StashCache access. This mode still relies on HTCondor's end-to-end credential management capabilities, for sending access tokens along with the jobs and refreshing tokens as needed.

This early OSG deployment experience has been especially helpful for working out SciTokens packaging and upgrade path details, since we have gone through upgrades of the SciTokens software (including file-server plug-ins) and credential profiles. Rolling out these updates across the distributed OSG infrastructure has required the SciTokens project to think about compatibility and versioning from the start.

\section{\label{sec:c}C/C++ Implementation}

In addition to our Java \cite{scitokens-java} and Python \cite{derek_weitzel_2018_1187173} SciTokens implementations, which we described in \cite{SciTokensPEARC18}, we have added a C/C++ implementation \cite{scitokenscpp}, which enables improved performance for our CVMFS and XrootD integrations.

It has also enabled development of a SciTokens Apache module, which implements an Apache user authentication type. The module is developed using the SciTokens authorization helper for CVMFS and the JWT Apache authentication module. In the module, the token is retrieved from the current request with the Apache Portable Runtime (APR)  library and verified with functions in the SciTokens C/C++ library.

The SciTokens C/C++ implementation has also enabled support for SciTokens as a native HTCondor authorization method (e.g., for authorizing access to the condor\_schedd). SciTokens C++ RPMs will soon be included in the OSG software distribution.

\section{\label{sec:oauth}OAuth Server Improvements}

We have updated our Token Server to support a more flexible policy language, including per-client policies based on SAML attributes and LDAP queries. We have also added support for OAuth token revocation \cite{RFC7009}, dynamic client registration \cite{RFC7591,RFC7592}, and mobile clients \cite{RFC8252}. 

\section{\label{sec:relatedwork}Related Work}

The SciTokens project has benefited from participation in the WLCG Authorization working group, including involvement in the development of a WLCG profile for JSON Web Tokens \cite{WLCGJWTProfile} that is compatible with SciTokens.

The SciTokens project has also engaged in interoperability testing with other JWT implementations in the scientific community, including INFN IAM\footnote{\url{https://iam.infn.it/}} and dCache\footnote{\url{https://www.dcache.org/}} services.

\section{\label{sec:conclusions}Conclusions}

The JSON Web Token and OAuth standards provide a solid foundation for distributed, capability-based authorization for scientific workflows. By enhancing existing components (CILogon, CVMFS, HTCondor, XrootD) to support the SciTokens model, we have provided a migration path from X.509 identity-based delegation to OAuth capability-based delegation for existing scientific infrastructures.

All SciTokens code is open source and published at \url{https://github.com/scitokens}.
The HTCondor CredMon is also open source, published at \url{https://github.com/htcondor/scitokens-credmon}. Visit \url{https://scitokens.org/} for the latest information about the SciTokens project.

\begin{acks}
This material is based upon work supported by the National Science Foundation under Grant No.~1738962.
\end{acks}

\bibliographystyle{ACM-Reference-Format}
\bibliography{scitokens}

\end{document}